\documentstyle[12pt,graphics,epsfig]{article}
\renewcommand{\theequation}{{\rm\thesection.\arabic{equation}}}
\newcommand{\be}{\begin{equation}}
\newcommand{\ee}{\end{equation}}
\newcommand{\f}{\begin{equation}}
\newcommand{\e}{\end{equation}}
\newcommand{\bea}{\begin{eqnarray}}
\newcommand{\eea}{\end{eqnarray}}

\newcommand{\D}{\displaystyle}

\begin{document}
\baselineskip 3.4ex
\title{Phonon-phonon interactions due to non-linear effects in a linear 
ion trap}
\author{Cyrille Marquet$^1$\footnote{Permanent address: MIP, Universit\'{e} 
Pierre et Marie Curie and D\'{e}partement de Physique, \'{E}cole Normale 
Sup\'{e}rieure, 75005 Paris, France.}, Ferdinand Schmidt-Kaler$^{2}$\\ and 
Daniel F.~V.  James$^1$\footnote{To whom correspondence should be addressed; 
TEL: +(505)-667-5436, FAX: +(505)-667-1931, email: dfvj@lanl.gov},\\
\small{$^1$ Theoretical Division T-4, Los Alamos National Laboratory,}\\
\small{Mail Stop B-283, P.O. Box 1663, Los Alamos, NM 87545, USA}\\
\small{$^2$ Institut f\"{u}r Experimentalphysik, Universit\"{a}t Innsbruck,}\\
\small{Technikerstra{\ss}e 25, A-6020 Innsbruck, Austria}}
\date{\today}
\maketitle
\begin{abstract} We examine in detail the theory of the intrinsic
non-linearities in the dynamics of trapped ions due to the Coulomb
interaction.  In particular the possibility of mode-mode coupling,
which can be a source of decoherence in trapped ion quantum computation,
or, alternatively, can be exploited for parametric down-conversion of
phonons, is discussed and conditions under which such coupling is
possible are derived.   \end{abstract}
\begin{center}
\bigskip
LA-UR-02-6745 \\
\bigskip
To be submitted to {\em Applied Physics B}.
\end{center}
\newpage
\section{Introduction}\setcounter{equation}{0}
Cold ions confined in electromagnetic traps and cooled by means of
lasers are a very important experimental system both for the study of
fundamental physics, such as cold non-neutral plasmas or quantum
dynamics, and for technological applications such as optical frequency
standards.  In the past few  years systems of this kind have been
the subject of intense study as a possible architecture for the
realization of a quantum computer \cite{CiracZoller,NistGate,MikeIke}
(for reviews of progress towards this goal, see \cite{NistReview,HotCold,
NistProg,SasBuz}).  In comparison with other experimental systems of
investigations of fundamental quantum phenomena, cold-trapped ions
offer a number of advantages.  In particular, it is a relatively
``clean'' system whose behavior is well characterized by theory;
simplified models of interactions between two-level systems
and quantum harmonic oscillators can be realized experimentally.
Considerable achievements in the production of non-classical states
of matter \cite{Nistfour} and other fundamental physics problems such as 
high-efficiency measurements of Bell's inequalities \cite{NistBell}
or tests of cavity quantum-electrodynamics \cite{InnsCQED}
have been made.

As is well known \cite{Earnshaw}, it is impossible to confine
charged particles by electrostatic forces alone.  To overcome
this problem, the radio-frequency Paul trap was developed: such
devices create an effective binding potential in the
$x$ and $y$ directions while a weaker static potential is applied in 
the $z$ direction \cite{Ghosh}.  When two or more ions are confined
in such a trap, they will repel each other due to the Coulomb force, 
resulting in confined charged plasmas with very low densities.
When sufficiently cold, the plasma will condense into a crystalline state.
In the highly anisotropic traps used for some atomic clocks \cite{hgclock}
and for quantum computing, this crystalline state 
is, for small enough numbers of ions, a simple chain of ions lying
in a straight line along the axis of weak binding.  In what follows, 
we shall be examining the oscillatory of ions confined in 
such a linear configuration.

The ions are assumed to be sufficiently cold that they undergo only small
oscillations around their equilibrium positions. In this case, the Lagrangian of
the motion can be expanded as a Taylor series around these
equilibrium positions.  As we shall see, the first term of this expansion
is a constant (which has no effect on the motion), 
the second term vanishes (by virtue of the condition for equilibrium);
the third term describes coupling of one ion's motion to all of the 
others (and can be easily resolved into a series of normal modes 
describing the ions' collective oscillations); the fourth term (whose
consequences are the main theme of this paper), describes an intrinsic 
coupling between the different modes. 

The effects of non-linearities in the coupling of harmonically bound 
particles is a problem with a long history: for example Fermi, Pasta
and Ulam investigated such dynamics computationally in 1954-55, using
the early Los Alamos MANIAC computer, with a view to explaining the 
equipartition of energy between modes as the system reaches thermal
equilibrium \cite{FPU}.  In the quantum realm, the non-linearities
lead to couplings between the normal modes of oscillation.
In certain ways this coupling is analogous to $\chi^{(2)}$ non-linear
optical effects \cite{BEZ}, in that, provided certain conditions are met, quanta
of one oscillatory mode can be down-converted into twinned pairs of
quanta of other modes. Obviously, since the Taylor expansion is an
infinite series, higher terms, describing more complex multi-mode
couplings, will be present; we will, however, not consider them in
this current work.

The paper is organized as follows. In Section 2 the Lagrangian of
the ion motion is presented and the derivation of the normal modes
of the ion oscillations is reprised; the classical 
description of mode-mode coupling in terms of the Hamiltonian, is 
formulated. In section 3 the properties of the mode-mode coupling
coefficients are discussed, and section 4 discusses the quantum theory
of this coupling and the  population transfers that it can cause.
Section 5 summarizes our conclusions.

\section{Lagrangian of the ion motion in the trap}
\setcounter{equation}{0}

\begin{figure}[th!]
\center{ \epsfig{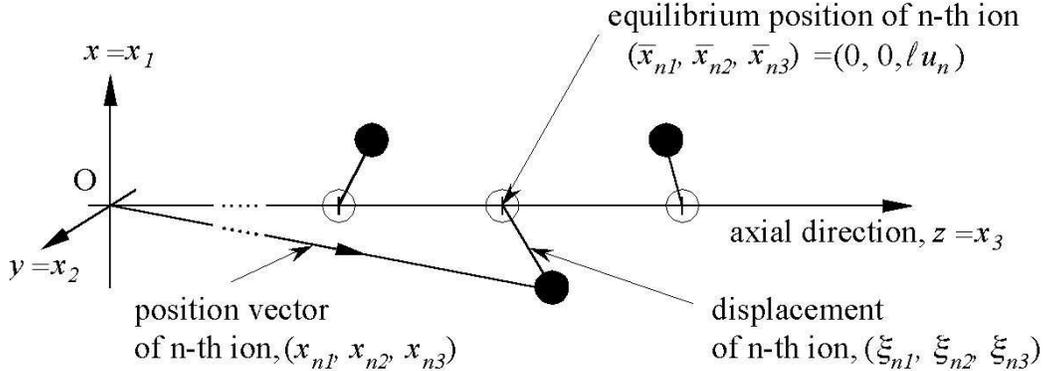}}
\caption{Illustrating the notation used in this paper.}
\label{geofig}
\end{figure}

Consider $N$ identical ions of mass $M$ and charge $Q$
confined in an effective three-dimensional harmonic potential
\footnote{In reality, the ions are experiencing a time varying force
whose effects can be modeled as an effective harmonic motion plus a
high-frequency oscillation called {\em micromotion}.  Experimental
techniques for minimizing this effect are explained in ref.\cite{mumo}.
Here we will assume that the amplitudes of transverse oscillations are
sufficiently small that micromotion can be neglected in this paper.}
. The
position of the $n$-th ion will be denoted $(x_{n1},x_{n2},x_{n3})$ 
(Fig.\ref{geofig}), 
where the ions are numbered in order of increasing value of their axial
positions $x_{n3}$, so that $n>m$ implies that $x_{n3}>x_{m3}$.
Besides the trap potential, each ion experiences a
Coulomb interaction with each of the other ions. Thus the total
potential energy of the ions is given by the following
expression:
\be 
V=\frac{M}{2}\sum_{n=1}^{N}\sum_{i=1}^{3}\omega_i^2\,x_{ni}^2
+\frac{Q^2}{8\pi{\epsilon}_0}
\sum_{\stackrel{\scriptstyle n,m=1}{m \neq n}}^N
\left[\sum_{i=1}^{3}\left(x_{ni}-x_{mi}\right)^{2}\right]^{-\frac{1}{2}}\:,
\ee
where $\omega_1$, $\omega_2$ and $\omega_3$ are the angular frequencies of
the trap in the three directions.   For simplicity, we will assume that the
trapping potentials are equal in  two transverse directions ($1$ and
$2$) and that the trapping potential in the axial direction ($3$)
is much weaker i.e.
\be
\omega_1=\omega_2=\omega_3/\sqrt{\alpha},
\ee
where $\alpha \ll 1$ is a dimensionless parameter characterizing the
anisotropy of the trap.

The equilibrium positions of the ions, denoted by $\bar{x}_{ni}$,
$(n=1,\ldots,N)$, $(i=1,2,3)$ are determined by the following equations:
\be
\left.\frac{{\partial}V}{{\partial}x_{ni}}\right|_0 =0\;\;\;(n=1,\ldots,N)\;\;\;(i=1,2,3),
\ee
where the subscript 0 denotes that the partial derivatives are evaluated 
at $x_{ni}=\bar{x}_{ni}$. The solutions of these equilibrium equations 
in strongly anisotropic trapping conditions (so that the ions are 
aligned along the weak axial direction $3$, i.e. $\bar{x}_{n1}=\bar{x}_{n2}=0$)
have been investigated by various authors \cite{steane,iontrapfun, 
Toddpaper} (see also \cite{Kielpinski,Morigi} for analysis of ion 
crystals with two different species of ion). It will be convenient
to use the dimensionless equilibrium positions defined by
$u_n=\bar{x}_{n3}/\ell$ where the length scale $\ell$ is defined 
by 
\be
\ell= \left( \frac{Q^2}{4 \pi \epsilon_0 M \omega_3^2} \right)^{1/3}.
\label{ldef}
\ee

The displacements of the ions from their equilibrium position 
are denoted $\xi_{ni}$, i.e. 
\be
x_{ni}(t)=\bar{x}_{ni}+\xi_{ni}(t).
\ee 
The Lagrangian describing the motion is then
\bea
L&=& T-V \nonumber\\
&=& \frac{M}{2}\sum_{n=1}^{N}\sum_{i=1}^{3}\dot{\xi}_{ni}^2 -
\frac{M}{2}\sum_{n=1}^{N}\sum_{i=1}^{3}\omega_i^2\,\left[\bar{x}_{ni}+\xi_{ni}(t)\right]^{2} \nonumber \\
&&-\frac{Q^2}{8\pi{\epsilon}_0}
\sum_{\stackrel{\scriptstyle n,m=1}{m \neq n}}^N
\left\{\sum_{i=1}^{3}\left[\bar{x}_{ni}+\xi_{ni}(t)-\bar{x}_{mi}-\xi_{mi}(t)\right]^{2}\right\}^{-\frac{1}{2}}.
\eea
Making a Taylor expansion about the equilibrium positions, this may be approximated as
\bea
L&=&\frac{M}{2}\sum_{n=1}^{N}\sum_{i=1}^{3}\dot{\xi}_{ni}^2-V_{0}
-\frac{1}{2}\sum_{m,n=1}^{N}\sum_{i,j=1}^{3}\left. \frac{\partial^2 V}{{\partial}x_{mi}{\partial}x_{nj}}\right|_0\xi_{mi}\xi_{nj}
\nonumber \\
&&-\frac{1}{6}\sum_{m,n,p=1}^{N}\sum_{i,j,k=1}^{3}
\left.\frac{\partial^3V}{{\partial}x_{mi}{\partial}x_{nj}{\partial}x_{p\,k}}\right|_0\xi_{mi}\xi_{nj}\xi_{p\,k}
+{\rm O}\left[\xi_{mi}^{4}\right].
\eea
Neglecting both the constant $V_{0}$ (which has no effect on the 
dynamics) and the higher order terms and evaluating the partial derivatives explicitly (see Appendix A), 
the Lagrangian may be approximated by following expression
\bea
L&\approx&
\frac{M}{2}\left[\sum_{n=1}^{N}\left(\dot{\xi}_{n3}\right)^2-
{\omega}_3^2\sum_{m,n=1}^{N}A_{mn}{\xi}_{m3}{\xi}_{n3}\right]\nonumber\\
&&+\frac{M}{2}\sum_{i=1}^{2}\left[\sum_{n=1}^{N}\left(\dot{\xi}_{ni}\right)^2-
{\omega}_3^2\sum_{m,n=1}^{N}B_{mn}{\xi}_{mi}{\xi}_{ni}\right]\nonumber\\
&&-\frac{M{\omega}_3^2}{2l}\sum_{m,n,p=1}^{N}C_{mnp}\,\xi_{p3}
\left(2\xi_{m3}\xi_{n3}-3\xi_{m1}\xi_{n1}-3\xi_{m2}\xi_{n2}\right).
\label{lag}
\eea
The first term represents the ions oscillations along the axial ($i=3$) 
direction, the second oscillations in the transverse directions 
($i=1,2$) and the third term represents coupling between these 
oscillations which are a direct and unavoidable consequence of the Coulomb 
interaction between the ions.  
The tensors $A_{mn}$, $B_{mn}$ and $C_{mnp}$ are given by
\bea 
A_{mn}&=&
\left\{
\begin{array}{lll}
	{\D 1+2\sum_{\stackrel{\scriptstyle p=1}{p \neq m}}^N 
	\frac{1}{\left|u_m-u_p\right|^3}} & \mbox{ if $m=n$ },\\
	\\
	{\D \frac{-2}{\left|u_m-u_n\right|^3}} & \mbox{ if $m\neq{n}$,
}\end{array}\right. \label{Adef}\\
\nonumber\\
B_{mn}&=&(\frac{1}{\alpha}+\frac{1}{2})\,{\delta}_{mn}-\frac{1}{2}A_{mn} ,
\label{Bdef}\\
\nonumber\\
C_{mnp}&=&\left\{
\begin{array}{lllll}
	{\D \sum_{\stackrel{\scriptstyle q=1}{q \neq m}}^N\frac{{\rm sgn}(q-m)}{(u_q-u_m)^4}}
	& \mbox{ if $m=n=p$ },\\
	\\
	{\D \frac{-{\rm sgn}(p-m)}{(u_p-u_m)^4}}& \mbox{ if $m=n\neq{p}$ },\\ 
	\\
	0 & \mbox{ if $m\neq{n\neq{p}}$,
}\end{array}\right. \label{Cdef}
\eea
where ${\rm sgn}(x)$ stands for the sign of $x$ and $\delta_{mn}$ is 
the Kronecker delta. The third rank tensor $C_{mnp}$ is symmetric 
under any exchange of two subscripts; thus, for example, 
$C_{224} = C_{242} =C_{422} =-1/(u_2-u_4)^4$ while 
$C_{442} = C_{244} =C_{424} = 1/(u_2-u_4)^4$.
All the elements of $A_{mn}$ and $C_{mnp}$ can be calculated
numerically from the dimensionless equilibrium positions of the ions.

The matrix $A_{mn}$ is real, symmetric and  positive definite. 
Thus its eigenvalues ${\mu}_p$ are non-negative.  The eigenvectors $b_n^{(p)}$
$(p=1,\ldots,N)$ are defined by the following formula:
\be
\sum_{n=1}^{N}A_{mn}b_n^{(p)}={\mu}_p\,b_m^{(p)}\;\;\;\;(p=1,\ldots,N),
\ee
where the eigenvectors are numbered in order
of increasing eigenvalue. The eigenvectors form a complete basis so that
\be
\sum_{p=1}^{N}b_m^{(p)}b_n^{(p\,)}={\delta}_{mn}\:,\hspace{2cm}
\sum_{m=1}^{N}b_m^{(p)}b_m^{(q)}={\delta}_{p\,q}.
\label{orthonorm}
\ee
For any value of $N$, the two first eigenvectors are
the center of mass mode $b^{(1)}=1/\sqrt{N}(1,1,\ldots,1)$, 
($\mu_1=1$) and the stretch mode $b^{(2)}=(1/{\cal N})$$(u_1,u_2,\ldots,u_N)$
($\mu_2=3$) where ${\cal N}=\sqrt{\sum_{n=1}^{N}u_n^2}$.
All the eigenvalues and eigenvectors can be calculated
numerically; their approximate values are given for example in
\cite{iontrapfun} for 2 to 10 ions. The dynamics of
these modes have been confirmed directly in experiments 
\cite{InnsModes}.
From eq.(\ref{Bdef}) we see that $B_{mn}$ has the 
same eigenvectors as $A_{mn}$ but different eigenvalues:
\be
\sum_{n=1}^{N}B_{mn} b_n^{(p)}=\gamma_p b_m^{(p)}\;\;\;\;(p=1,\ldots,N),
\ee
where the eigenvalue $\gamma_p$ is related to the longitudinal 
eigenvalue by the formula
\be
\gamma_p=\frac{1}{\alpha}+\frac{1}{2}-\frac{\mu_p}{2}.
\label{defgamma}
\ee
If $\alpha > \alpha_{crit} = 2/(\mu_N -1)$ 
the matrix $B_{nm}$ is not positive definite; this
implies unstable transverse oscillation modes,
and results in the formation of ``zig-zag'' crystal structures
 \cite{zigzag}. We shall assume however that $\alpha$ is kept 
sufficiently small that this situation does not arise.
Note that, with this convention, the transverse oscillation modes
are numbered in order of {\em decreasing} eigenvalue, so that
the $p=1$ center of mass mode has the largest eigenvalue.

Using these eigenvectors, the normal modes of
the ions' oscillations in the three directions are defined as follows: 
\bea
X_{p}(t)&=&\sum_{n=1}^{N}b_n^{(p)}\xi_{n1}(t)\:\:,\nonumber\\
Y_{p}(t)&=&\sum_{n=1}^{N}b_n^{(p)}\xi_{n2}(t)\:\:,\\
Z_{p}(t)&=&\sum_{n=1}^{N}b_n^{(p)}\xi_{n3}(t)\:\:.\nonumber
\eea
The Lagrangian eq.(\ref{lag}) can be rewritten in terms of these 
normal modes quite straightforwardly.  Not surprisingly, one finds that, 
neglecting
the fourth order term, the Lagrangian becomes a sum of Lagrangians of harmonic 
oscillators.  The canonical momenta conjugate to $X_n$ is $\Pi_n^x=\partial 
L/\partial\dot X_n$ (with analogous definitions for $\Pi_n^y$ and 
$\Pi_n^z$).  Using these momenta, the Hamiltonian for the ion motion is
\be 
H=H_0+H_I,
\ee 
where $H_0$ is the Hamiltonian for all of the uncoupled collective oscillations, 
i.e.
\bea 
H_0&=&\sum_{n=1}^{N}\left[\frac{{\Pi_n^x}^2}{2M}+\frac{M\Omega_n^2X_n^2}{2}\right] \nonumber\\
&+&\sum_{n=1}^{N}\left[\frac{{\Pi_n^y}^2}{2M}+\frac{M\Omega_n^2Y_n^2}{2}\right]\\
&+&\sum_{n=1}^{N}\left[\frac{{\Pi_n^z}^2}{2M}+\frac{M\nu_n^2Z_n^2}{2}\right],\nonumber
\label{classH0}
\eea
$\nu_n=\omega_3\sqrt{\mu_n}$ being the angular frequency of the $n$-th mode in the z
direction and $\Omega_n=\omega_3\sqrt{\gamma_n}$ the angular frequency
of the $n$-th transverse mode. 
The Hamiltonian $H_I$ describes the perturbation which, in certain 
circumstances, couples these various modes.
It is given by
\be
H_I=\frac{M{\omega}_z^2}{2l}\sum_{m,n,p=1}^{N}D_{mnp}Z_p(2Z_mZ_n-3X_mX_n-3Y_mY_n),
\label{classH1}
\ee
where the mode-mode coupling coefficients $D_{pqr}$ are defined by
\be
D_{pqr}=\sum_{l,m,n=1}^{N}C_{lmn}b_l^{(p)}b_m^{(q)}b_n^{(r)}.
\label{Ddef}
\ee
Note that the sign of $D_{pqr}$ depends on the sign of the 
eigenvector ${\bf b}^{(p)}$, which is undetermined.  We will adopt 
the convention that the sign of the $N$-th component of ${\bf b}^{(p)}$
is always positive, for all modes and values of $N$.

We should mention that other mechanisms by which resonant interactions
between ion motion in different directions are coupled have been 
studied.  In particular Werth {\em et al.} at the University of Mainz 
have investigated experimentally the parametric coupling of different
stable configurations of molecular ion clouds \cite{Werth}.  
Cross-mode couplings due to static field imperfections, which can be 
a cause of heating or of decoherence, were also discussed in ref.\cite{NistReview} 
sec 4.1.8.

\section{Properties of the Coupling Coefficients}\setcounter{equation}{0}

In this section we discuss some of the properties of the coefficients
$C_{mnp}$ and  $D_{mnp}$, which are central to the subsequent 
development of phonon-phonon interactions in ion traps.
The first important symmetry property stems directly from the definition of 
$C_{mnp}$, eq.(\ref{Cdef}): The tensor $C_{mnp}$ is symmetric under 
exchange of two subscripts; further, because of the definition (\ref{Ddef}),
so is $D_{mnp}$.  From the definition of $C_{mnp}$ eq.(\ref{Cdef}), we also
have the property $C_{mmn}=-C_{nnm}$ (provided that $m\neq n$).
These properties imply that
\be
\sum_{p=1}^{N}C_{mnp}=0. \label{artichoke}
\ee
This can be shown as follows:  if $m\neq n$, we have  
$\sum_{p=1}^{N}C_{mnp} = C_{mnn}+C_{mnm}$, since $C_{mnp}$ is
always zero if all three indices are different.  However, since
$C_{mnm}= C_{mmn}= -C_{nnm}= -C_{mnn}$, the sum is zero.  If
$m = n$, $\sum_{p=1}^{N}C_{mmp} = C_{mmm}+\sum_{p=1, p\neq 
m}^{N}C_{mmp}$, which, by the definition of $C_{mmm}$, 
eq.(\ref{Cdef}),  must also be zero.
The identity eq.(\ref{artichoke}) implies that, since
$b^{(1)} =(1/\sqrt{N})\{1,1,\ldots1\}$,
\be
D_{mn1}\equiv \frac{1}{\sqrt{N}}\sum_{i,j=1}^{N}b_i^{(m)}b_j^{(n)}\sum_{k=1}^{N}C_{ijk}=0.
\label{cauliflower}
\ee
Thus, when considering the three-mode mixing described by eq.(\ref{classH1})
the center of mass $(p=1)$ mode has zero coupling strength to 
any other mode.  Physically, this can be explained as follows.  The 
$p=1$ mode is special since it is the only mode in which the center of 
mass of the crystal is displaced: as can be seen from 
eq.(\ref{orthonorm}), $\sum^{N}_{m=1}b^{(p)}_{m}=0$ if $p\neq 1$.  
Thus to excite or de-excite the $p=1$ mode requires the application of
an external force to change the momentum of the crystal as a whole, 
while no such force is required to transfer energy between any of the other 
modes.

This result has important consequences: firstly, if one 
wanted to avoid mode-mode coupling entirely, for example when performing 
quantum logic operations between ions, it would behoove one to use the 
center of mass mode as the ``quantum bus'' as originally proposed by 
Cirac and Zoller \cite{CiracZoller}.  However, in some experiments 
\cite{Turchette}, it is 
found that the quantum state of the center of mass mode is rapidly 
degraded by excitations from various extraneous sources.  
Our result eq.(\ref{cauliflower}) tends to rule out coupling to other
ion oscillation modes which may have been imperfectly cooled as a sources
of such heating.  Furthermore, when the center of mass mode is heated 
preferentially, it will not cause heating of the other modes. 

A second interesting property of  $C_{mnp}$ can be expressed as 
follows:
\be
\sum_{p=1}^{N}u_p\,C_{mnp}=\frac{1}{2}\left[\delta_{mn}-A_{mn}\right],
\label{cabage}
\ee
where $A_{mn}$ is the ion coupling tensor defined by eq.(\ref{Adef}).
This result can be demonstrated as follows: if $m\neq n$, then  
$\sum_{p=1}^{N}C_{mnp}u_{p} = C_{mnn}u_{n}+C_{mnm}u_{m} = 
(u_{n}-u_{m}) C_{mnn} = 1/|u_{n}-u_{m}|^{3}$, where the last step 
results from the definition of $C_{mnn}$ in terms of the scaled equilibrium 
positions $u_{p}$.  For the case $m = n$, $\sum_{p=1}^{N}C_{mnp}u_{p} =
-u_{m}\sum_{p=1, p\neq m}^{N}C_{mmp}+\sum_{p=1, p\neq m}^{N}C_{mmp} 
u_{p} = -\sum_{p=1, p\neq m}^{N}/|u_{p}-u_{m}|^{3}$.  Comparing these 
expressions involving the $u_{p}$'s with the definition of $A_{mn}$, 
eq.(\ref{Adef}), we obtain eq.(\ref{cabage}).
This result implies the following
\be
D_{mn2}
=\sum_{i,j=1}^{N}b_i^{(m)}b_j^{(n)}\sum_{k=1}^{N}\frac{u_k}{\cal N}\,C_{ijk}
=\frac{1-\mu_m}{2{\cal N}}\delta_{mn},
\ee
where, as before, ${\cal N}=\sqrt{\sum_{n=1}^{N}u_n^2}$.  Thus
the coupling of the first stretch mode ($p$=2) is 
constrained so that it will {\em only} be coupled to a single 
other mode rather than to two distinct modes.

For $N=2$ and $N=3$, we can determine the coefficients $D_{mnp}$
algebraically, using the exact expressions for the equilibrium 
positions and mode vectors which can be obtained in those two simple 
cases.  We find that
\bea
N=2:D_{222}&=&-2^{\frac{1}{6}}, \nonumber \\
N=3:D_{233}&=&-\frac{3}{\sqrt{2}}\left(\frac{4}{5}\right)^{\frac{4}{3}}, 
\vspace{5mm}
D_{222}=-\frac{1}{\sqrt{2}}\left(\frac{4}{5}\right)^{\frac{1}{3}}.
\eea
All the other coefficients must be determined numerically, although 
this is a reasonably straightforward task. The approximate numerical 
values of the non-zero coefficients for 2 to 10 ions are given in Tables 
2a and 2b (Appendix B).

\section{Quantum motion of the ions}\setcounter{equation}{0}
\subsection{Quantization of the Hamiltonian}

We can now consider the quantum motion of the ions by introducing the 
following position and momentum operators 
\bea
Z_n(t)\rightarrow\hat{Z}_n =\sqrt{\frac{\hbar}{2M\nu_n}}(\hat{a}_n+\hat{a}_n^{\dagger})
&&
\Pi_n^z(t)\rightarrow\hat{\Pi}_n^z=
i\sqrt{\frac{M\hbar\,\nu_n}{2}}(\hat{a}_n^{\dagger}-\hat{a}_n),\nonumber \\
X_n(t)\rightarrow\hat{X}_n=\sqrt{\frac{\hbar}{2M\Omega_n}}(\hat{b}_n+\hat{b}_n^{\dagger})
&&
\Pi_n^x(t)\rightarrow\hat{\Pi}_n^x=
i\sqrt{\frac{M\hbar\,\Omega_n}{2}}(\hat{b}_n^{\dagger}-\hat{b}_n),\nonumber \\
Y_n(t)\rightarrow\hat{Y}_n=\sqrt{\frac{\hbar}{2M\Omega_n}}(\hat{c}_n+\hat{c}_n^{\dagger})
&&
\Pi_n^y(t)\rightarrow\hat{\Pi}_n^y=
i\sqrt{\frac{M\hbar\,\Omega_n}{2}}(\hat{c}_n^{\dagger}-\hat{c}_n),\nonumber \\
\eea
where, as before, the $Z$-coordinate refers to motion along the axial 
direction of the trap, while the $X$- and $Y$-coordinates correspond 
to displacements in the transverse direction.  These operators obey
the canonical commutation relations
\bea
\left[\hat{X}_p,\hat{\Pi}_q^x\right]= 
\left[\hat{Y}_p,\hat{\Pi}_p^y\right]=\left[\hat{Z}_p,\hat{\Pi}_p^z\right]&=&
i\hbar\delta_{p,q}, \nonumber \\
\left[\hat{a}_p,\hat{a}_q^{\dagger}\right]=
\left[\hat{b}_p,\hat{b}_q^{\dagger}\right]=
\left[\hat{c}_p,\hat{c}_q^{\dagger}\right]&=&
\delta_{p,q}. \label{comrel}
\eea
The Hamiltonian becomes:
\be
\hat{H}=\hat{H}_0+\hat{H}_I,
\ee
where
\bea
\hat{H}_0&=&\sum_{n=1}^{N}\hbar 
\left\{\nu_n \hat{a}_n^{\dagger}\hat{a}_n +
\Omega_n\left[\hat{b}_n^{\dagger}\hat{b}_n + \hat{c}_n^{\dagger}\hat{c}_n\right]
\right\} \nonumber \\
\hat{H}_I&=&
\varepsilon \hbar \omega_{3}
\sum_{m,n,p=1}^{N}\frac{D_{mnp}}{\sqrt[4]{\mu_p}}(\hat{a}_p+
\hat{a}_p^{\dagger})
\left(\frac{2}{\sqrt[4]{\mu_m\mu_n}}(\hat{a}_m+
\hat{a}_m^{\dagger})(\hat{a}_n+\hat{a}_n^{\dagger})\right.\nonumber\\
&&\left.-\frac{3}{\sqrt[4]{\gamma_m\gamma_n}}
\left[(\hat{b}_m+\hat{b}_m^{\dagger})(\hat{b}_n+\hat{b}_n^{\dagger})
+(\hat{c}_m+\hat{c}_m^{\dagger})(\hat{c}_n+\hat{c}_n^{\dagger})\right]\right)
\label{montagu}
\eea
where $\varepsilon$ is a  dimensionless quantity which 
characterizes the strength of the 
non-linearity; it is given by
\begin{equation} 
\varepsilon = \frac{1}{4 \sqrt{2}} \left[\frac{\hbar 
\omega_{3}}{\alpha_{\rm{fsc}}^{2} M c^{2}}  \right]^{1/6} ,
\end{equation}
where $\alpha_{\rm{fsc}}$ is the fine structure constant.
Alternatively,  $\varepsilon= \sigma /4 \ell$, 
where $\sigma = \sqrt{\hbar/2 M \omega_{3}}$ is the approximate width of
the wavepackets of the individual ions, and $\ell$, defined by 
eq.(\ref{ldef}), is the length scale of the order of the ions' 
separation.
 divided by the ions' separation.  Since
both $D_{mnp}$ and $\mu_p$ are, at most, of the order of unity, while
$\gamma_p$ is assumed to be much larger than unity, we have:
\be
\left|\frac{\hat{H}_I}{\hat{H}_0}\right| \sim \varepsilon.
\ee
Values of $\varepsilon$ for a variety of ions and traps are given in 
Table \ref{epstab}.  These show this parameter tends to have a small value in
experimental circumstances, implying that the treatment of $\hat{H}_I$ as a
perturbation can be justified.

\begin{table}[h]
\begin{center}
\begin{tabular}{|l|l|l|l|}
\hline
Ion                        &    $\omega_{3}$         &    $\varepsilon$          &    $\varepsilon\omega_{3}$\\
\hline \hline
$^9 \mbox{\rm Be}^{+}$     &   ($2 \pi $) 5.0  MHz   &    1.06 $\times 10^{-3}$  &     $(2\pi)$ 5.30 kHz      \\
$^{40} \mbox{\rm Ca}^{+}$  &   ($2 \pi $) 2.0  MHz   &    7.09 $\times 10^{-4}$  &     $(2\pi)$ 1.42 kHz      \\
$^{88} \mbox{\rm Sr}^{+}$  &   ($2 \pi $) 200  kHz   &    4.24 $\times 10^{-4}$  &     $(2\pi)$ 85 Hz         \\
$^{112} \mbox{\rm Cd}^{+}$ &   ($2 \pi $) 2.8  MHz   &    6.32 $\times 10^{-4}$  &     $(2\pi)$ 1.77 kHz      \\
\hline
\end{tabular}
\end{center}
\caption{Values of the parameter $\varepsilon$ for various 
ions and trapping frequencies.}
\label{epstab}
\end{table}

We will denote the eigenstates and eigenvalues of $\hat{H}_0$ as:
\bea
|\Psi\rangle&=&|\left\{n_1^x,\ldots,n_N^x 
\right\},\left\{n_1^y,\ldots,n_N^y\right\},\left\{n_1^z,\ldots,n_N^z\right\}\rangle \label{christie}
\\
E_\Psi&=&\hbar\sum_{p=1}^{N}\Omega_p\left(n_p^x+n_p^y\right)+\nu_p n_p^z.
\eea

\subsection{Resonance conditions}
Fermi's golden rule \cite{FGR} implies that population
transfer between two eigenstates $|i\rangle$ and $|f\rangle$ of
$\hat{H}_0$ induced by the interaction $\hat{H}_I$ 
can only occur if the matrix element ${\langle}f|\hat{H}_I|i\rangle$ 
is non-zero and if the energies of  $|i\rangle$ and $|f\rangle$ are equal 
\footnote{This is equivalent to the energy matching condition
in parametric down-conversion.  Since here the modes do {\em not} represent 
travelling waves, there is no momentum conservation (or phase matching) 
condition as there is the non-linear optics.}
.  These conditions allow us to neglect a considerable number of
the terms that occur when eq.(\ref{montagu}) is  expanded.
For example, because the square roots of the eigenvalues of $A_{m,n}$ are all
irrational numbers (with the sole exception of $\mu_{1}$), it follows
that $\pm\left(\sqrt{\mu_p}\pm\sqrt{\mu_m}\pm\sqrt{\mu_n}\right)\neq{0}$;
hence we can neglect all the terms of the form
$(\hat{a}_p+\hat{a}_p^{\dagger})(\hat{a}_m+\hat{a}_m^{\dagger})(\hat{a}_n+\hat{a}_n^{\dagger})$.
Similarly terms involving three creation or three annihilation 
operators
can be ignored.  Making these approximations, and using the symmetry 
properties of $D_{mnp}$ and the
commutation relations (\ref{comrel}), we obtain the following simplified
expression for the interaction Hamiltonian, (in the interaction picture)
\bea
\hat{H}_I &\approx&
-3 \varepsilon \hbar \omega_{3}
\sum_{m,n,p=1}^{N}\frac{D_{mnp}}{\sqrt[4]{\gamma_m\gamma_n\mu_p}}
\left[ 
2 \hat{a}_p 
\left(\hat{b}_m^{\dagger}\hat{b}_n+\hat{c}_m^{\dagger}\hat{c}_n 
\right) e^{i \Delta^{(-)}_{mnp} \omega_{z} t} \right.\nonumber \\
&&+\hat{a}_p 
\left.\left(\hat{b}_m^{\dagger}\hat{b}_n^{\dagger}+\hat{c}_m^{\dagger}\hat{c}_n^{\dagger} 
\right) e^{i \Delta^{(+)}_{mnp} \omega_{z} t }
\right] +h.a., \label{bernardo}
\eea
where $h.a.$ stands for the Hermitian adjoint of the preceding term, 
and $\Delta^{(\pm)}_{mnp} = \sqrt{\gamma_{m}} \pm \sqrt{\gamma_{n}} - \sqrt{\mu_{p}}$.
We shall refer to the term proportional to $\exp(i \Delta^{(-)}_{mnp} 
\omega_{z} t)$, which involves the creation of a transverse phonon and
simultaneous annihilation of both a longitudinal and a transverse phonon
(and the reverse process, contained within the hermitin adjoint part 
of eq.(\ref{bernardo})), as a
resonance of the first kind; the term with $\exp(i \Delta^{(+)}_{mnp} 
\omega_{z} t)$, (which creates two transverse phonons in different
modes by annihilating a longitudinal phonon) will be called a 
resonance of the second kind.

\begin{figure}[th!]
\center{ \epsfig{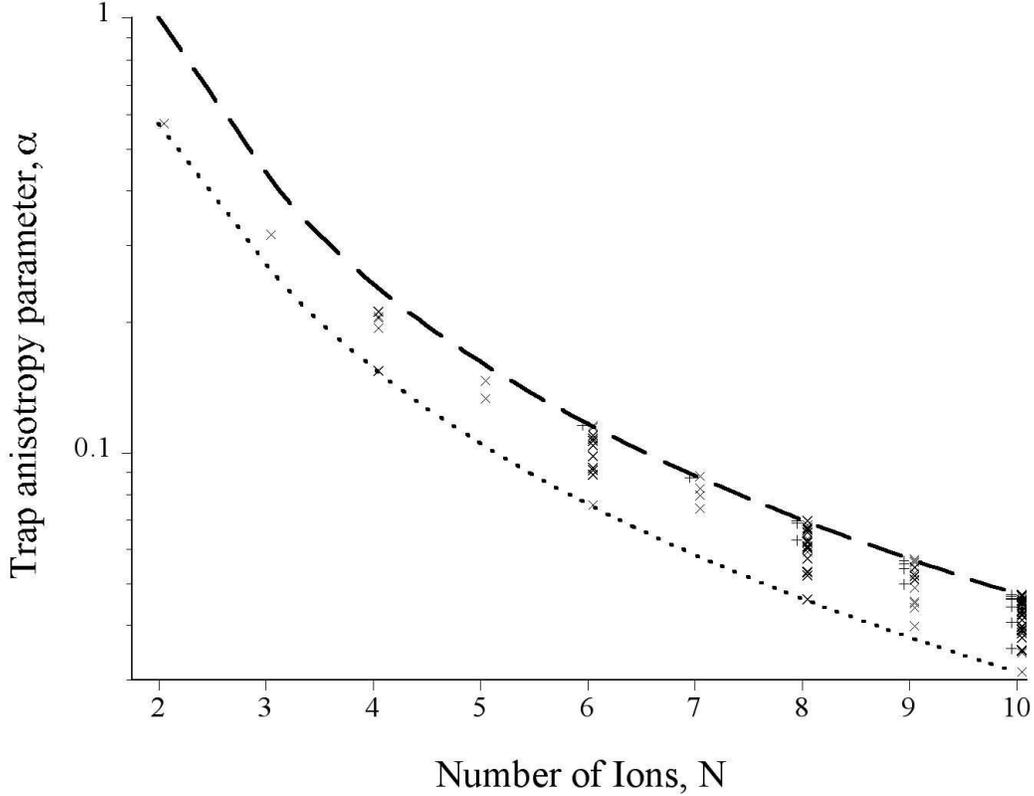}}
\caption{Values of the anisotropy parameter $\alpha$ for
which resonant three-mode mixing can occur.  The horizontal/verticle 
crosses represent resonances of the first kind ($\Delta^{(-)}_{mnp} =0$) and the 
diagonal crosses represent resonances of the second kind ($\Delta^{(+)}_{mnp} =0$). 
The upper limiting curve represents the critical value 
$\alpha_{crit}$, above which the linear confirguration of the ions becomes unstable.
The lower curve represents the highest value of $\alpha$ for which 
resonant mode-mode coupling cannot occur (see eq.(4.11)).}
\label{alphaplot}
\end{figure}

Since the only experimentally controllable parameter in
the definitions of $\Delta^{(\pm)}_{mnp}$ 
is the anisotropy parameter $\alpha$,  it
is natural to ask at what values of $\alpha$ a resonance can occur.
A necessary condition for resonance (i.e. for $\Delta^{(+)}_{mnp} = 
0$ {\em or} $\Delta^{(-)}_{mnp} = 0$) is as follows:
\be
\alpha=\frac{16\mu_{p}}{4\mu_{p}^2+\mu_{m}^2+\mu_{n}^2-8\mu_{p}+4\mu_{p}\mu_{m}+
4\mu_{p}\mu_{n}-2\mu_{m}\mu_{n}}. \label{alphamnp}
\ee
Using the known values of the eigenavalues $\mu_{p}$ 
(ref.\cite{iontrapfun}, table 2), appropriate values for $\alpha$ can
be found straightforwardly.  
However, not all of these values of $\alpha$ given by (\ref{alphamnp})
correspond to a resonance, 
and one must determine whether or not the appropriate
condition for either type of resonance is satisfied by direct 
substitution into the formulas for $\Delta^{(\pm)}_{mnp}$.  
Since $\sqrt{\mu_{p}}\neq 0$, resonances of the first kind are only 
possible when two distinct transverse modes are involved (i.e. $m\neq 
n$); resonances of the second kind can occur involving a single 
transverse mode (i.e. $m$ can be equal to $n$). 
Resonances of the first kind tend to be very weak; the coupling 
coefficients do not exceed $1.6 \times 10^{-3}$ for $L\leq 10$, and are always
zero  for $L \leq 6$.
The values of
$\alpha$ for the two types of resonance are given in Tables 2a and 2b 
(Appendix B) and are 
plotted for different
numbers of ions in figure \ref{alphaplot}.
Only those values less than the critical value 
$\alpha_{crit}$ are included, since our analysis is based on the  
assumption that the ions are in a linear configuration.

One important result stems from this analysis.  Resonance {\em cannot}
occur for values of $\alpha$ below the minimum value given by
\be
{\everymath {\displaystyle}
\alpha_{min} = \left\{
\begin{array}{ll}
\frac{16 \mu_{N}}{\mu_{N}\left(9 \mu_{N}+2 
\mu_{N-1}-8\right)+\mu_{N-1}^{2}} > \frac{4}{3 \mu_{N}-2} & \mbox{if 
$N$ is odd},\\
\\
\frac{4}{3 \mu_{N} -2} & \mbox{if $N$ is even}.\\
\end{array}\right.}
\ee
Thus, should one wish to avoid any form of mode-mode coupling due to
the forth order terms we are considering in this paper, this can be 
ensured by using a trap with an anisotropy parameter 
smaller than $4/(3 \mu_{N} -2)$.  This can also be expressed in terms 
of the critical anisoptry for onset of zig-zag, $\alpha_{crit}$: there
will be no three-mode mixing if $\alpha<4 
\alpha_{crit}/(\alpha_{crit}+6)$.  

\subsection{Example of mode-mode coupling}
As a specific example of the above general analysis, let us assume
that we have six ions confined in a trap with anisotropy factor
$\alpha$=0.09151.  For this value of $\alpha$, a resonance of
the second kind ($\Delta^{(+)}_{mnp}=0$) occurs for the three
modes $(m,n,p)=(5,6,5)$ and also
for $(m,n,p)=(6,5,5)$ (since there is symmetry between $m$ and $n$ for
resonances of the second kind).

The coupling Hamiltonian can be simplified by neglecting
all off-resonant terms; for this case we get
\bea
\hat{H}_I&\simeq&
-3 \varepsilon \hbar \omega_{3} \frac{D_{565}}{\sqrt[4]{\mu_{5}\gamma_{5}\gamma_{6}}}
\hat{a}_{5}\left(\hat{b}_{5}^{\dagger}\hat{b}_{6}^{\dagger}+\hat{c}_{5}^{\dagger}\hat{c}_{6}^{\dagger}\right)\nonumber \\
&&+
-3 \varepsilon \hbar \omega_{3} 
\frac{D_{655}}{\sqrt[4]{\mu_{5}\gamma_{6}\gamma_{5}}}
\hat{a}_{5}\left(\hat{b}_{6}^{\dagger}\hat{b}_{5}^{\dagger}+\hat{c}_{6}^{\dagger}\hat{c}_{5}^{\dagger}\right) +h.a.\nonumber \\
&=&
- \varepsilon \hbar \omega_{3} \frac{6 D_{565}}{\sqrt[4]{\mu_{5}\gamma_{5}\gamma_{6}}}
\hat{a}_{5}\left(\hat{b}_{5}^{\dagger}\hat{b}_{6}^{\dagger}+\hat{c}_{5}^{\dagger}\hat{c}_{6}^{\dagger}\right) +h.a., \label{heath}
\eea
where we have used the fact that $D_{655}=D_{565}$ and that 
$\left[\hat{b}_{5}^{\dagger}, \hat{b}_{6}^{\dagger}\right]=0$ (and 
similarly for the $\hat{c}_{m}^{\dagger}$).  From Table 2 we see
that $D_{655}$=4.2528 and  $\alpha$=0.09151; from Table 2 of 
ref.\cite{iontrapfun} we find $\mu_{5}$ = 13.51; using 
eq.(\ref{defgamma}) we find $\gamma_{5}$= 4.6709 and $\gamma_{6}$=2.2949;
thus $6 D_{655}/\sqrt[4]{\mu_{5}\gamma_{5}\gamma_{6}}$ = 7.3556.

\begin{figure}[ht]
\center{ \epsfig{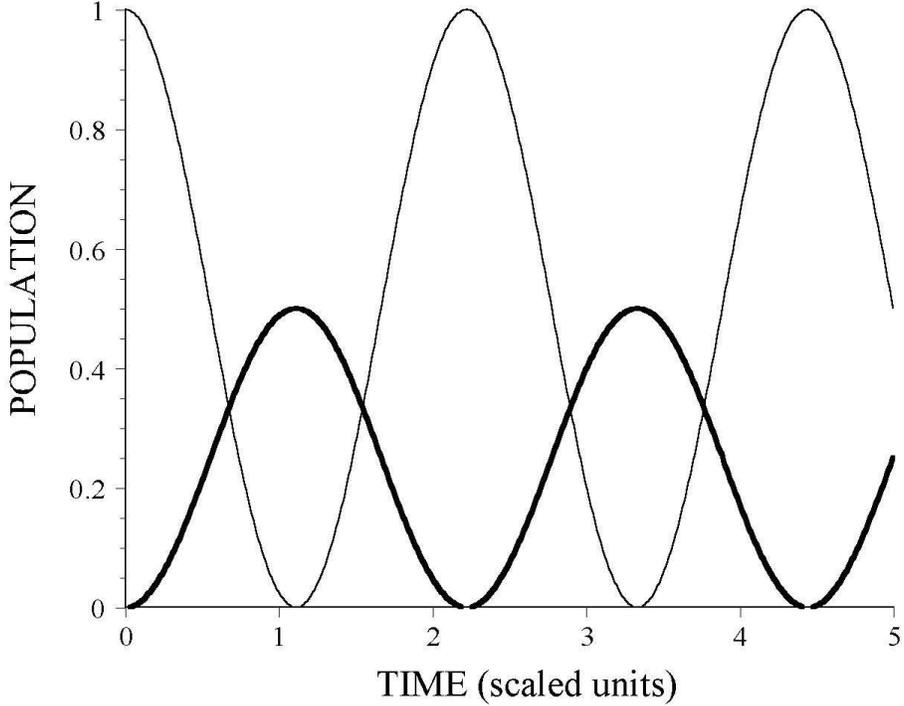}}
\caption{The evolution of
the populations when the initial state is $|\psi\rangle$. The thin
line represents $\left|\psi(t)\right|^2$ and the wide line
$\left|\phi(t)\right|^2=\left|\chi(t)\right|^2$.  Time is in units
of $1/\Gamma$.}
\label{scharnhorst}
\end{figure}

\begin{figure}[ht]
\center{ \epsfig{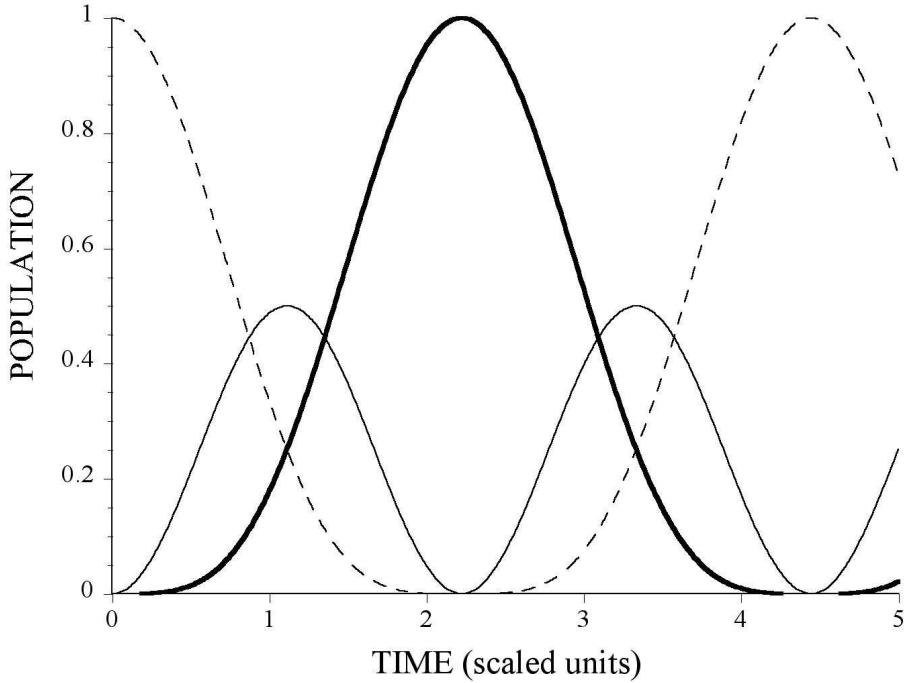}}
\caption{The evolution of
the populations when the initial state is $|\phi\rangle$. The thin
line represents $\left|\psi(t)\right|^2$, the dashed line
$\left|\phi(t)\right|^2$ and the wide line
$\left|\chi(t)\right|^2$.  Time is in units
of $1/\Gamma$.}
\label{bismarck}
\end{figure}

As an example we will consider the case of coupling between 
three states defined by
\bea
|\Psi\rangle &=&|\left\{0,0,0,0,0,0\right\},\left\{0,0,0,0,0,0\right\},\left\{0,0,0,0,1,0\right\}\rangle, \nonumber\\
|\Phi\rangle &=&|\left\{0,0,0,0,0,0\right\},\left\{0,0,0,0,1,1\right\},\left\{0,0,0,0,0,0\right\}\rangle, \\
|\chi\rangle &=&|\left\{0,0,0,0,1,1\right\},\left\{0,0,0,0,0,0\right\},\left\{0,0,0,0,0,0\right\}\rangle, \nonumber
\eea
where we have used the notation for the motional quantum states of the ion modes defined in eq.(\ref{christie}).
Note that the coupling defined by eq.(\ref{heath}) only involves the 
5th longitudinal mode and the 5th and 6th transverse modes; the 
states of the other modes are unaffected; thus although we have set 
these other modes to have zero population in the definitions of 
$|\psi\rangle$, $|\phi\rangle$ and $|\chi\rangle$, their state can be 
arbitrary without changing the analysis.  The interaction picture state describing the
system is given by:
\be
|\Psi(t)\rangle=\psi(t)|\Psi\rangle+\phi(t)|\Phi\rangle+\chi(t)|\chi\rangle.
\ee
Since, due to the resonance condition, $\left[\hat{H}_0,\hat{H}_I\right]=0$ and the interaction picture
coupling Hamiltonian is:
\be
e^{\frac{i}{\hbar}\hat{H}_0t}\hat{H}_I e^{-\frac{i}{\hbar}\hat{H}_0t}=\hat{H}_I.
\ee
Thus the Schr\"{o}dinger equation describing the evolution of the 
system is given by
\be
i\hbar\frac{d}{dt}|\Psi_I(t)\rangle=\hat{H}_I|\Psi_I(t)\rangle,
\ee
which implies that the probability amplitudes obey the following set 
of coupled equations:
\bea
i\dot{\psi}(t)&=&-\Gamma\left[\phi(t)+\chi(t)\right], \nonumber \\
i\dot{\phi}(t)&=&-\Gamma\,\psi(t), \\
i\dot{\chi}(t)&=&-\Gamma\,\psi(t),\nonumber
\eea
where $\Gamma= \varepsilon\omega_{3} (6 D_{565} /\sqrt[4]{\mu_{5}\gamma_{5}\gamma_{6}})$.
The solution of these equations can be obtained straightforwardly 
using Laplace transforms:
\bea
\psi(t)&=&\psi(0)\cos\left(\sqrt{2}\Gamma{t}\right)+\frac{i\left(\phi(0)+
\chi(0)\right)}{\sqrt{2}}\sin\left(\sqrt{2}\Gamma{t}\right), \nonumber \\
\phi(t)&=&\frac{i\psi(0)}{\sqrt{2}}\sin\left(\sqrt{2}\Gamma{t}\right)+
\phi(0)\cos^2\left(\frac{\Gamma{t}}{\sqrt{2}}\right)-\chi(0)\sin^2\left(\frac{\Gamma{t}}{\sqrt{2}}\right), \\
\chi(t)&=&\frac{i\psi(0)}{\sqrt{2}}\sin\left(\sqrt{2}\Gamma{t}\right)-
\phi(0)\sin^2\left(\frac{\Gamma{t}}{\sqrt{2}}\right)+\chi(0)\cos^2\left(\frac{\Gamma{t}}{\sqrt{2}}\right). \nonumber
\eea 
These solutions are plotted in Figure \ref{scharnhorst} for the case
$\psi(0)=1$ (i.e. the longitudinal phonon mode initially excited) and
in Figure \ref{bismarck} for the case
$\phi(0)=1$ (i.e. two transverse modes in the x-direction initially
excited).

\newpage
\subsection{Entanglement}
Let us consider in detail the first example, shown in Fig.3. The 
crystal of six ions was prepared with a single quanta in the fifth
oscillatory mode, i.e. the state
\be
|\Psi(0)\rangle =|\left\{0,0,0,0,0,0\right\},\left\{0,0,0,0,0,0\right\},\left\{0,0,0,0,1,0\right\}\rangle.
\ee
After a time $t_{1}=\pi/2 \sqrt{2}\Gamma$ the crystal will have evolved into the
state 
\bea
|\Psi(t_{1})\rangle &=& \frac{i}{\sqrt{2}} 
\left(|\left\{0,0,0,0,0,0\right\},\left\{0,0,0,0,1,1\right\},\left\{0,0,0,0,0,0\right\}\rangle \right. \nonumber\\
&&+\left.|\left\{0,0,0,0,1,1\right\},\left\{0,0,0,0,0,0\right\},\left\{0,0,0,0,0,0\right\}\rangle\right).
\eea 
This later state has the property of being entangled: it cannot be 
written as a product of the state of 5-th and of the 6-th oscillatory 
modes.  This can be thought of as analogous to the type of 
polarization entangled states of photon pairs generated in optical 
down-conversion experiments \cite{Paul}.  The two transverse directions
of oscillation are analogous to the photon polarizations, while the 
oscillatory mode is analogous to the spatial mode of the photons. The
initially excited longitudinal oscillations play the role of the pump 
laser. Controlling this process in ion traps will be achieved by 
switching the trap anisotropy parameter $\alpha$ into the appropriate 
resonance parameter for the appropriate amount of time, then rapidly 
switching it to a non-resonant value:  this will have the effect of 
turning on the phonon-phonon interaction for a specific time.  
Furthermore, the technology of preparing oscillatory quantum states of trapped 
ions is at the moment more advanced than current state of the art for preparing
quantum states of the electromagnetic field.  Specifically, one can 
hope to prepare the ``pump'' longitudinal mode in
a single excitation Fock state, while creating single photons is still 
somewhat problematic.  Thus in principle, deterministic preparation of 
oscillatory entangled states may be attained, in contrast to the 
non-deterministic, post-selection state preparation in quantum optics 
experiments.  Furthermore the method of entangled oscillatory state
preparation described here, 
combined with the variety of techniques for creating entanglement of 
the internal degrees of freedom raises the possibility of creating 
{\em hyper-entangled} states in ion traps \cite{hype}.

\section{Conclusion}\setcounter{equation}{0}
In this paper we have analyzed the dynamics of ions confined in 
harmonic traps.  The normal modes of the ions' oscillations 
are intrinsically coupled by the Coulomb interaction.  This 
``three-mode mixing'' is in some ways analogous to non-linear optical 
effects.  There are two important results which our analysis has 
revealed: first, if one wishes to avoid these effects altogether, the 
trap anisotropy, characterized by the parameter $\alpha$ defined by 
eq.(2.2), must be less than a certain value given by eq.(4.11).  
Secondly, this effect in principle may be exploited to create 
entangled motional states of the ions, by carefully controlling the
ions' dynamics, as outline in section 4.3 and 4.4.  
How the resulting entangled states may be exploited for tasks in quantum 
information is a promising avenue for further investigation

\section*{Acknowledgments}
The authors would like to thank Dana Berkeland and Eddy Timmermans
for useful discussions, Sara Schneider and John Grondalski for 
reading and commenting on the manuscript and Albert Petschek and William Beyer for 
drawing our attention to reference \cite{FPU}. CM would also like to
thank Los Alamos National Laboratory for its hospitality during his
visit and ENS for providing travel funds.  This work was supported
in part by the Los Alamos National Laboratory LDRD program.

\section*{Appendix A: Derivation of Eqs.(\ref{Adef}), (\ref{Bdef}) and 
(\ref{Cdef}).}
\renewcommand{\theequation}{{\rm A.\arabic{equation}}}\setcounter{equation}{0}

In this
Appendix, we show how to derive the Lagrangian eq.(2.8) from the
eq.(2.7). We will use the following notation: 
\bea
R_i^{(mn)}&=&x_{mi}-x_{ni},\\
R^{(mn)}&=&\left[\sum_{i=1}^{3}\left(x_{mi}-x_{ni}\right)^2\right]^{\frac{1}{2}}.
\eea
The potential energy of the ions is 
\be
V=\frac{M}{2}\sum_{n=1}^{N}\sum_{i=1}^{3}\omega_i^2\,x_{ni}^2+\frac{Z^2e^2}{8\pi{\epsilon}_0}
\sum_{\stackrel{\scriptstyle n,m=1}{m \neq n}}^N \frac{1}{R^{(mn)}}.
\ee 
Thus we have:
\bea
\frac{\partial{V}}{\partial{x_{mi}}}&=&M\omega_i^2\left(x_{mi}-\frac{\ell^3\omega_3^2}{\omega_i^2}
\sum_{\stackrel{\scriptstyle p=1}{p \neq m}}^N
\frac{R_i^{(mp)}}{{R^{(mp)}}^3}\right),\\
\frac{\partial^2{V}}{\partial{x_{mi}}\partial{x_{nj}}}&=&M\omega_i^2\left[\delta_{mn}\delta_{ij}-
\frac{\ell^3\omega_3^2}{\omega_i^2}
\sum_{\stackrel{\scriptstyle p=1}{p \neq m}}^N
\frac{\delta_{mn}-\delta_{pn}}{{R^{(mp)}}^3}
\left(\delta_{ij}-3\frac{R_i^{(mp)}R_j^{(mp)}}{{R^{(mp)}}^2}\right)\right], 
\nonumber \\
&&\\
\frac{\partial^3{V}}{\partial{x_{mi}}\partial{x_{nj}}\partial{x_{pk}}}&=&
3M\ell^3\omega_3^2 \sum_{\stackrel{\scriptstyle q=1}{q \neq m}}^N
(\delta_{mn}-\delta_{qn})(\delta_{mp}-\delta_{qp})\nonumber\\
&&\times\left(\frac{\delta_{ij}R_k^{(mq)}+\delta_{ik}R_j^{(mq)}+\delta_{jk}R_i^{(mq)}}{{R^{(mq)}}^5}
-5\frac{R_i^{(mq)}R_j^{(mq)}R_k^{(mq)}}{{R^{(mq)}}^7}\right). 
\nonumber\\
\eea
These terms can be evaluated at equilibrium by making the following 
substitutions:
\bea
\bar{R}_i^{(mn)}&=&\ell\,\delta_{i3}(u_m-u_n), \\
\bar{R}^{(mn)}&=&\ell\,|u_m-u_n|.
\eea 
We thus obtain:
\bea
\left.\frac{\partial{V}}{\partial{x_{mi}}}\right|_0 &=& M\omega_i^2\delta_{i3}\left(u_{m}-
\sum_{\stackrel{\scriptstyle n=1}{n \neq m}}^N
\frac{sgn(u_m-u_n)}{(u_m-u_n)^2}\right) \label{larry}, \\
\left.\frac{\partial^2{V}}{\partial{x_{mi}}\partial{x_{nj}}}\right|_0&=&M\omega_3^2\left[\delta_{mn}\delta_{ij}\frac{\omega_i^2}{\omega_3^2}-
\sum_{\stackrel{\scriptstyle p=1}{p \neq m}}^N
\frac{\delta_{mn}-\delta_{pn}}{|u_m-u_p|^3}
\left(\delta_{ij}-3\delta_{i3}\delta_{j3}\right)\right], \nonumber \\
&&\label{curly}\\
\left.\frac{\partial^3{V}}{\partial{x_{mi}}\partial{x_{nj}}\partial{x_{pk}}}\right|_0&=&
\frac{3M\omega_3^2}{\ell}
\sum_{\stackrel{\scriptstyle q=1}{q \neq m}}^N
\frac{sgn(u_m-u_q)(\delta_{mn}-\delta_{qn})(\delta_{mp}-\delta_{qp})}{(u_m-u_q)^4}\hspace{1cm}\nonumber\\
&&\times\left(\delta_{ij}\delta_{k3}+\delta_{ik}\delta_{j3}+\delta_{jk}\delta_{i3}-5\delta_{i3}\delta_{j3}\delta_{k3}\right). \label{moe}
\eea
Equation (\ref{larry}) give the equations that determine the 
equilibrium positions $u_m$.
From eq.(\ref{curly}) we can write
\be
\frac{1}{2}\sum_{m,n=1}^{N}\sum_{i,j=1}^{3}\left. \frac{\partial^2 V}{{\partial}x_{mi}{\partial}x_{nj}}\right|_0\xi_{mi}\xi_{nj}
=\frac{M\omega_3^2}{2}\sum_{m,n=1}^{N}A_{mn}\xi_{m3}\xi_{n3}+B_{mn}\left(\xi_{m1}\xi_{n1}+\xi_{m2}\xi_{n2}\right),
\ee
with 
\be
A_{mn}=\delta_{mn}+2
\sum_{\stackrel{\scriptstyle p=1}{p \neq m}}^N
\frac{\delta_{mn}-\delta_{pn}}{|u_m-u_p|^3},
\ee
and 
\be
B_{mn}=\frac{\delta_{mn}}{\alpha}-
\sum_{\stackrel{\scriptstyle p=1}{p \neq m}}^N
\frac{\delta_{mn}-\delta_{pn}}{|u_m-u_p|^3},
\ee
which are equivalent to the expressions given in Section 2. 
And, finally:
\bea
\frac{1}{6}&&\sum_{m,n,p=1}^{N}\sum_{i,j,k=1}^{3}
\left.\frac{\partial^3V}{{\partial}x_{mi}{\partial}x_{nj}{\partial}x_{p\,k}}\right|_0\xi_{mi}\xi_{nj}\xi_{p\,k}
\nonumber \\
&&=\frac{M\omega_3^2}{2\,\ell}\sum_{mnp=1}^{N}C_{mnp}\,\xi_{p3}\left(2\xi_{m3}\xi_{n3}-3\xi_{m1}\xi_{n1}-3\xi_{m2}\xi_{n2}\right),
\eea
where
\be
C_{mnp}=
\sum_{\stackrel{\scriptstyle q=1}{q \neq m}}^N
\frac{sgn(u_q-u_m)(\delta_{mn}-\delta_{qn})(\delta_{mp}-\delta_{qp})}{(u_m-u_q)^4}.
\ee

\section*{Appendix B}
\renewcommand{\theequation}{{\rm B.\arabic{equation}}}\setcounter{equation}{0}
The following tables list the non-zero values of the mode cross-coupling 
coefficients $D_{mnp}$ defined by eq.(\ref{Ddef}) for crystals of 2 
to 10 ions, together with the associated resonant anisotropy 
parameter $\alpha$.  Other values may be determined via the symmetry relation 
$D_{mnp}=D_{mpn}=D_{npm}=D_{nmp}=D_{pmn}=D_{pnm}$.  Note however that 
the value of $\alpha$ is symmetric only under interchange of the 
first two indices.  \\
\vspace{5mm}

\noindent
{\bf Table 2a}: Coupling constants for resonances of the first kind
(in which a transverse phonon creates a longitudinal phonon and a transverse 
phonon).
\scriptsize
\begin{center}
\begin{tabular}{| lll | lll | lll |}
\hline
\{m,n,p\}     &   $D_{mnp}$      & $\alpha_{res}$ &   \{m,n,p\}     &   $D_{mnp}$      &$\alpha_{res}$ &   \{m,n,p\}     &   $D_{mnp}$      &$\alpha_{res}$\\
\hline
\multicolumn{3}{|l|} {N=6} &\multicolumn{3}{|l|} {N=7} &\multicolumn{3}{|l|} {} \\
\hline
\{3,6,3\}     &    2.8395e-4     &    0.11575     &  \{4,7,3\}      &    5.4794e-4     &    0.087700   &                 &                  &            \\
\hline
\multicolumn{9}{|l|} {N=8}\\
\hline
\{3,8,3\}     &    3.1152e-06    &    0.062943    &    \{4,8,4\}    &    1.0031e-3     &    0.069797   &   \{5,8,3\}     &    7.4469e-4     &    0.068879\\
\hline
\multicolumn{9}{|l|} {N=9}\\
\hline
\{3,9,4\}     &    5.7509e-06    &    0.054162    &    \{4,9,3\}    &    5.7509e-06    &    0.049960   &   \{5,9,4\}     &    1.2967e-3     &    0.056583\\
\{6,9,3\}     &    8.8502e-4     &    0.055637    &                 &                  &               &                 &                  &            \\
\hline
\multicolumn{9}{|l|} {N=10}\\
\hline
\{3,10,3\}    &    5.9416e-08    &    0.035465    &    \{3,10,5\}   &    7.3591e-06    &    0.046313   &   \{4,10,4\}    &    1.0248e-05    &    0.044211\\
\{5,10,3\}    &    7.3591e-06    &    0.040715    &   \{5,10,5\}    &    1.5984e-3     &    0.047291   &   \{6,10,4\}    &    1.4703e-3     &    0.046822\\
\{7,10,3\}    &    9.8451e-4     &    0.045964    &                 &                  &               &                 &                  &            \\
\hline
\end{tabular}
\end{center}

\newpage
\normalsize
\noindent
{\bf Table 2b}: Resonances of the second kind,
(in which a longitudinal phonon creates two transverse phonons). 
\scriptsize

\begin{center}
\begin{tabular}{| lll | lll | lll |}
\hline
\{m,n,p\}    &   $D_{mnp}$   &$\alpha_{res}$ &   \{m,n,p\}     &   $D_{mnp}$  &$\alpha_{res}$ &   \{m,n,p\}     &   $D_{mnp}$   &   $\alpha_{res}$  \\
\hline
\multicolumn{3}{|l|} {N=2} &\multicolumn{6}{|l|} {N=3}  \\
\hline
\{2,2,2\}    &    -1.1225    &    0.57143    &    \{3,2,3\}    &    -1.5754   &    0.30917    &    \{3,3,2\}    &    -1.5754    &    0.31746\\
\hline
\multicolumn{9}{|l|} {N=4}\\
\hline
\{3,3,4\}    &    -1.8332    &    0.21132    &    \{4,2,4\}    &    -1.9493   &    0.19337   &   \{4,3,3\}    &    -1.8332    &    0.20560 \\
\{4,4,2\}    &    -1.9493    &    0.20391    &    \{4,4,4\}    &    2.1084    &    0.15429   &                &               &            \\
\hline
\multicolumn{9}{|l|} {N=5}\\
\hline
\{4,3,5\}    &    -1.9683    &    0.14895    &    \{4,4,4\}    &    -0.89355  &    0.15387   &   \{5,2,5\}    &    -2.2887   &    0.13340\\
\{5,3,4\}    &    -1.9683    &    0.14187    &    \{5,4,3\}    &    -1.9683   &    0.14619   &   \{5,4,5\}    &    3.1611    &    0.11561\\
\{5,5,2\}    &    -2.2887    &    0.14311    &    \{5,5,4\}    &    3.1611    &    0.11668   &                &              &           \\
\hline
\multicolumn{9}{|l|} {N=6}\\
\hline
\{4,4,6\}    &    -1.8850    &    0.11439    &    \{5,3,6\}    &    -2.0625   &    0.10985   &   \{5,4,5\}    &   8.1287e-2  &    0.11527 \\
\{5,5,6\}    &    4.2528     &    0.092393   &    \{6,2,6\}    &    -2.6084   &    0.098229  &    \{6,3,5\}   &    -2.0625   &    0.10398 \\
\{6,4,4\}    &    -1.8850    &    0.10784    &    \{6,4,6\}    &    4.0185    &    0.088947  &   \{6,5,3\}    &    -2.0625   &    0.10937 \\
\{6,5,5\}    &    4.2528     &    0.091510   &   \{6,6,2\}     &    -2.6084   &    0.10658   &    \{6,6,4\}   &    4.0185    &    0.091151\\
\{6,6,6\}    &    -2.3706    &    0.075762   &                 &              &              &                &              &            \\
\hline
\multicolumn{9}{|l|} {N=7}\\
\hline
\{6,3,7\}    &    -2.1382    &    0.084291   &    \{6,5,7\}   &    4.9336    &    0.074398   &   \{6,6,6\}    &    4.3077    &    0.075526 \\
\{7,2,7\}    &    -2.9149    &    0.075728   &   \{7,3,4\}    &    5.4794e-4 &    0.088234   &    \{7,3,6\}   &    -2.1382   &    0.079773\\
\{7,4,5\}    &    -1.7957    &    0.082749   &    \{7,4,7\}   &    4.7748    &    0.070376   &   \{7,5,4\}    &    -1.7957   &    0.084569\\
\{7,5,6\}    &    4.9336     &    0.072836   &   \{7,6,3\}    &    -2.1382   &    0.085059   &    \{7,6,5\}   &    4.9336    &    0.074000\\
\{7,6,7\}    &    -3.8751    &    0.062565   &    \{7,7,2\}   &    -2.9149   &    0.082796   &   \{7,7,4\}    &    4.7748    &    0.073152\\
\{7,7,6\}    &    -3.8751    &    0.062861   &&&&&&\\
\hline
\multicolumn{9}{|l|} {N=8}\\
\hline
\{6,6,8\}    &    5.2871    &    0.062147   &    \{7,3,8\}   &    -2.2035   &    0.066792   &   \{7,5,8\}    &    5.4173    &    0.060792\\
\{7,6,7\}    &    3.6190    &    0.062378   &   \{7,7,6\}    &    3.6190    &    0.062674   &    \{7,7,8\}   &    -5.8640   &    0.053277\\
\{8,2,8\}    &    -3.2119   &    0.060393   &    \{8,3,5\}   &    7.4469e-4 &    0.069615   &   \{8,3,7\}    &    -2.2035   &    0.063341\\
\{8,4,6\}    &    -1.7158   &    0.065623   &   \{8,4,8\}    &    5.4691    &    0.057068   &    \{8,5,5\}   &    -1.5767   &    0.067200\\
\{8,5,7\}    &    5.4173    &    0.059134   &    \{8,6,4\}   &    -1.7158   &    0.068081   &   \{8,6,6\}    &    5.2871    &    0.060526\\
\{8,6,8\}    &    -5.1946   &    0.052159   &   \{8,7,3\}    &    -2.2035   &    0.068170   &    \{8,7,5\}   &    5.4173    &    0.061000\\
\{8,7,7\}    &    -5.8640   &    0.053018   &    \{8,8,2\}   &    -3.2119   &    0.066389   &   \{8,8,4\}    &    5.4691    &    0.060029\\
\{8,8,6\}    &    -5.1946   &    0.052896   &   \{8,8,8\}    &    2.1526    &    0.046043   &                &              &            \\
\hline

\multicolumn{9}{|l|} {N=9}\\
\hline
\{7,6,9\}    &    5.4146    &    0.052246   &    \{7,7,8\}    &    1.6006    &    0.053125    &    \{8,3,9\}    &    -2.2623   &    0.054311\\
\{8,5,9\}    &    5.7898    &    0.050462   &    \{8,6,8\}    &    2.6715    &    0.052000    &    \{8,7,7\}    &    1.6006    &    0.052867\\   
\{8,7,9\}    &    -7.3423   &    0.045500   &    \{8,8,6\}    &    2.6715    &    0.052744    &    \{8,8,8\}    &    -7.5531   &    0.045900\\
\{9,2,9\}    &    -3.5017   &    0.049433   &    \{9,3,6\}    &    8.8502e-4 &    0.056425    &    \{9,3,8\}    &    -2.2623   &    0.051650\\    
\{9,4,5\}    &   1.2967e-3  &    0.056883   &    \{9,4,7\}    &    -1.6468   &    0.053425    &    \{9,4,9\}    &    6.1214    &    0.047245\\ 
\{9,5,6\}    &    -1.4030   &    0.054734   &    \{9,5,8\}    &    5.7898    &    0.048927    &    \{9,6,5\}    &    -1.4030   &    0.055586\\
\{9,6,9\}    &    -6.3927   &    0.044000   &    \{9,7,4\}    &    -1.6468   &    0.056011    &    \{9,7,6\}    &    5.4146    &    0.050996\\
\{9,7,8\}    &    -7.3423   &    0.044970   &    \{9,8,3\}    &   -2.2623    &    0.055954    &    \{9,8,5\}    &    5.7898    &    0.051126\\
\{9,8,7\}    &    -7.3423   &    0.045400   &    \{9,8,9\}    &    3.8083    &    0.039874    &    \{9,9,2\}    &    -3.5017   &    0.054558\\
\{9,9,4\}    &    6.1214    &    0.050180   &    \{9,9,6\}    &    -6.3927   &    0.045081    &    \{9,9,8\}    &    3.8083    &    0.039988\\ 
\hline
\multicolumn{9}{|l|} {N=10}\\
\hline
\{7,7,10\}   &    5.2107    &    0.044976    &    \{8,6,10\}   &    5.4385    &    0.044332    &    \{8,7,9\}    &    -0.42626   &    0.045380    \\
\{8,8,8\}    &    -1.9896   &    0.045779    &    \{8,8,10\}   &    -8.6478   &    0.039673    &    \{9,3,10\}   &    -2.3165    &    0.045099\\
\{9,4,9\}    &    3.3150    &    0.047113    &    \{9,5,10\}   &    6.0930    &    0.042512    &    \{9,6,9\}    &    1.6336     &    0.043878\\
\{9,7,8\}    &    -0.42626  &    0.044846    &    \{9,7,10\}   &    -8.5012   &    0.039123    &    \{9,8,7\}    &    -0.42626   &    0.045279\\
\{9,8,9\}    &    -8.0166   &    0.039760    &    \{9,9,6\}    &    1.6336    &    0.044957    &    \{9,9,8\}    &    -8.0166    &    0.039874 \\
\{9,9,10\}   &    6.3437    &    0.035161    &    \{10,2,10\}  &    -3.7855   &    0.041300    &    \{10,3,7\}   &    9.8451 e-04&    0.046733\\
\{10,3,9\}   &    -2.3165   &    0.043017    &    \{10,4,6\}   &    1.4703e-3 &    0.047276    &    \{10,4,8\}   &    -1.5874    &    0.044422\\
\{10,4,10\}  &    6.7435    &    0.039798    &    \{10,5,7\}   &    -1.2642   &    0.045500    &    \{10,5,7\}   &    -1.2642    &    0.045500\\
\{10,5,9\}   &    6.0930    &    0.041163    &    \{10,6,6\}   &    -1.1721   &    0.046259    &    \{10,6,8\}   &    5.4385     &    0.042265\\
\{10,6,10\}  &    -7.5034   &    0.037564    &    \{10,7,5\}   &    -1.2642   &    0.046721    &    \{10,7,7\}   &    5.2107     &    0.043067\\
\{10,7,9\}   &    -8.5012   &    0.038483    &    \{10,8,4\}   &    -1.5874   &    0.046923    &    \{10,8,6\}   &    5.4385     &    0.043500\\
\{10,8,8\}   &    -8.6478   &    0.039085    &    \{10,8,10\}  &    5.3952    &    0.034687    &    \{10,9,3\}   &    -2.3165    &    0.046823\\
\{10,9,5\}   &    6.0930    &    0.043468    &    \{10,9,7\}   &    -8.5012   &    0.039269    &    \{10,9,9\}   &    6.3437     &    0.035057\\
\{10,10,2\}  &    -3.7855   &    0.045724    &    \{10,10,4\}  &    6.7435    &    0.042604    &    \{10,10,6\}  &    -7.5034    &    0.038860\\
\{10,10,8\}  &    5.3952    &    0.035000    &    \{10,10,10\} &    -1.7362   &    0.031318    &                 &               &            \\
\hline
\end{tabular}
\end{center}

\end{document}